\documentclass[conference]{IEEEtran}
\usepackage{cite}

\usepackage[pdftex]{graphicx}
\usepackage{subfig}
\usepackage{smartdiagram}
\usepackage{forest}
\usepackage{amsmath}
\usepackage{hyperref}
\usepackage{url}

\begin{document}

\title{Efficient Data Management in Neutron Scattering Data Reduction Workflows at ORNL}

\author{
\IEEEauthorblockN{William F Godoy}
\IEEEauthorblockA{
\textit{Computer Science and}\\
\textit{Mathematics Division,}\\
\textit{Oak Ridge National Laboratory}\\
Oak Ridge, TN, USA\\
Email: godoywf@ornl.gov}
\and
\IEEEauthorblockN{Peter F Peterson}
\IEEEauthorblockA{
\textit{Computer Science and}\\
\textit{Mathematics Division,}\\
\textit{Oak Ridge National Laboratory}\\
Oak Ridge, TN, USA\\
Email: petersonpf@ornl.gov}
\and
\IEEEauthorblockN{Steven E Hahn}
\IEEEauthorblockA{
\textit{Computer Science and}\\
\textit{Mathematics Division,}\\
\textit{Oak Ridge National Laboratory}\\
Oak Ridge, TN, USA\\
Email: hahnse@ornl.gov}
\and
\IEEEauthorblockN{Jay J Billings}
\IEEEauthorblockA{
\textit{Computer Science and}\\
\textit{Mathematics Division,}\\
\textit{Oak Ridge National Laboratory}\\
Email: billingsjj@ornl.gov}
}
\maketitle

\begin{abstract}
Oak Ridge National Laboratory (ORNL) experimental neutron science facilities produce 1.2\,TB a day of raw event-based data that is stored using the standard metadata-rich NeXus schema built on top of the HDF5 file format. Performance of several data reduction workflows is largely determined by the amount of time spent on the loading and processing algorithms in Mantid, an open-source data analysis framework used across several neutron sciences facilities around the world.
The present work introduces new data management algorithms to address identified input output (I/O) bottlenecks on Mantid. First, we introduce an in-memory binary-tree metadata index that resemble NeXus data access patterns to provide a scalable search and extraction mechanism. Second, data encapsulation in Mantid algorithms is optimally redesigned to reduce the total compute and memory runtime footprint associated with metadata I/O reconstruction tasks.
Results from this work show speed ups in wall-clock time on ORNL data reduction workflows, ranging from 11\% to 30\% depending on the complexity of the targeted instrument-specific data. Nevertheless, we highlight the need for more research to address reduction challenges as experimental data volumes increase.
\end{abstract}


\section*{}
This manuscript has been authored by UT-Battelle, LLC under Contract No. DEAC05-00OR22725 with the U.S. Department of Energy. The United States Government retains and the publisher, by accepting the article for publication, acknowledges that the United States Government retains a nonexclusive, paid-up, irrevocable, world-wide license to publish or reproduce the published form of this manuscript, or allow others to do so, for United States Government purposes. The Department of Energy will provide public access to these results of federally sponsored research in accordance with the DOE Public Access Plan (\url{http://energy.gov/downloads/doe-public-access-plan}).

\IEEEpeerreviewmaketitle

\section{Introduction}
Vast quantities of data are produced at two of the largest neutron source facilities in the world hosted at Oak Ridge National Laboratory (ORNL): the High Flux Isotope Reactor (HFIR) and the Spallation Neutron Source (SNS)~\cite{Neutrons}. Neutron scattering data produced at ORNL is used to address major scientific challenges across several industries. Currently ORNL instruments produce experimental data at a rate of 1.2\,TB a day, for a grand total of 1.6\,PB, with plans to expand the current volumes as new instruments, for example the VENUS beamline~\cite{Neutrons,BILHEUX201555}, become available.

Instruments at ORNL's HFIR and SNS facilities record individual neutron events~\cite{Granroth:ut5002} containing three essential elements: i) detector pixel identifier, ii) total neutrons' time-of-flight (TOF) from source to detector, and iii) wall-clock time of the proton pulse the neutron is associated with~\cite{PETERSON201524}. The vast amount of raw event data is stored using the metadata-rich standard NeXus schema~\cite{Konnecke:po5029}, built on top of the self-describing HDF5 hierarchical data file format~\cite{hdf5}. Each instrument at HFIR and SNS stores a subset of the NeXus schema according to its application. This data is hosted at ORNL computing facilities and available to users via remote access for their scientific needs ~\cite{Campbell_2010}.

\begin{figure}[!t]
\centering
\includegraphics[width=3in]{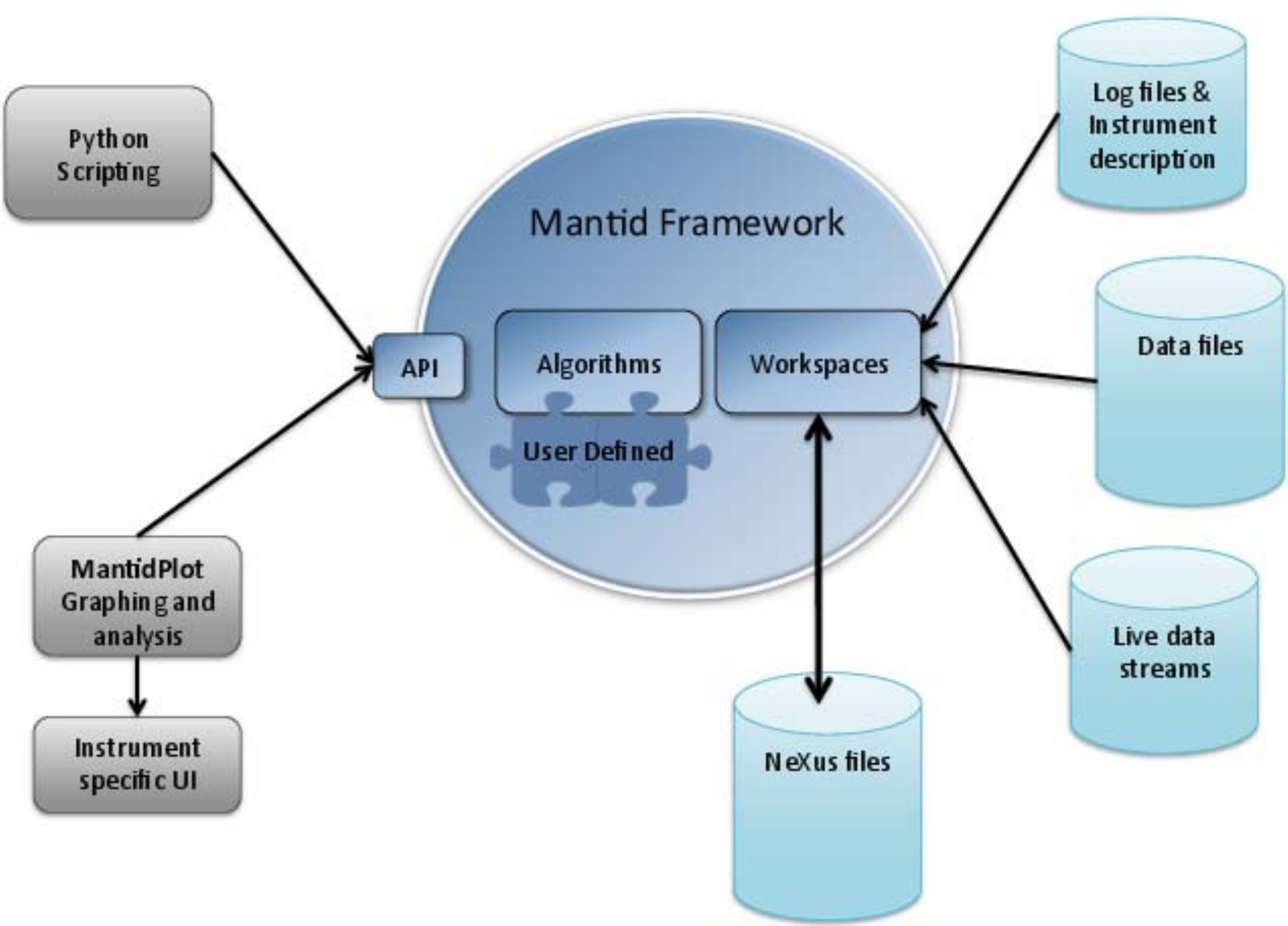}
\caption{Overview of the central role of the Mantid framework in data reduction workflows, from~\cite{ARNOLD2014156}. The step going from NeXus files to (event) Workspaces has been identified as a bottleneck at ORNL.}
\label{fig:Mantid}
\end{figure}

As shown in Fig.~\ref{fig:Mantid}, the stored NeXus datasets are loaded for post-processing by several data reduction workflows using the open-source data analysis and visualization Mantid framework~\cite{ARNOLD2014156}, written in C++~\cite{10.5555/2543987}. Mantid is part of an international collaboration between several neutron science facilities around the world; including ORNL's SNS and HFIR, the ISIS Neutron and Muons Source~\cite{THOMASON201961}, and The Institut Laue–Langevin (ILL)~\cite{AGERON1989197}. Loading NeXus files is an essential component in existing production data reduction workflows deployed to the facilities users. Mantid creates an in-memory data structure named an ``event workspace" to interpret raw event data by loading and processing algorithm operations on NeXus files. The latter operation has been identified as a major bottleneck in data reduction workflows~\cite{6972268}. Tackling these data I/O bottlenecks is critical for integrating novel paradigms, such as machine learning algorithms on vast amounts of experimental data, as they become more important across neutron science applications~\cite{9005968,8752943}.

As presented by Foster ~\textit{et al.}~\cite{8287761}, data I/O bottlenecks have been largely identified in high performance computing (HPC) co-design data reduction efforts. Furthermore, Alam~\textit{et al.}~\cite{10.1145/2159352.2159356} refers to the ``metadata wall" as one of the critical aspects in the performance of parallel file systems as more rich self-describing data is produced. Current efforts have been deployed in HPC systems to address metadata and data related bottlenecks at scale, such as ADIOS 2~\cite{GODOY2020100561} and ExaHDF5~\cite{SurenByna145}. As described by Zhang~\textit{et al.}~\cite{8990452}, proper metadata indexing is essential for efficient search and information discovery as scientific applications continue to produce large amounts of data. Much of the data generated from experiments, observations, and simulations is stored using self-describing data formats; in which the metadata and data can be accessed efficiently all at once. However, the authors also argue that very few systematic studies exist on the discovery of in-memory index strategies for different scientific applications. Diederich and Milton~\cite{109104} proposed the creation of metadata structures that are domain-specific, as opposed to well-established knowledge-based data models. The latter metadata problem has been identified in this work as one of the bottlenecks to address in the targeted neutron scattering data reduction workflows.

The present work introduces optimal data management strategies to address current I/O bottlenecks in the Mantid framework processing stages of NeXus datasets. In particular, managing the metadata entries in-memory to reduce the compute and memory runtime footprint. First, a suitable memory-persistent  binary-tree~\cite{10.1145/1734663.1734671} structure is introduced to speed up search and extraction operations by using an ``absolute-path" metadata entry index that matches processing operations on Mantid. The goal is to replace the current hierarchical approach to reconstruct indices using a ``relative-path" approach similar to walking through the directories of a file system. This incurs in added cost as memory and disk input output (I/O) resources are used ``on-demand". Second, Mantid's algorithms architecture encapsulation is reformulated to facilitate persistent data sharing across stages of processing NeXus files. These changes in the architecture allow for reusable information, thus reducing current bottlenecks associate with computing and memory run time footprint.

The remainder of the article is organized as follows. Section~\ref{sec:event_data} describes the NeXus format used for the raw event-based neutron data stored at ORNL SNS/HFIR facilities and a description of the current data reduction operations and challenges in the Mantid framework. Section~\ref{sec:proposed} presents the proposed 
data management strategies in Mantid: introduction of an in-memory binary-tree metadata indexing structure and modifying the encapsulation on Mantid algoritms.
The impact is shown in section~\ref{sec:results} illustrating the consistent speed ups obtained with the proposed strategy for different SNS and HFIR instruments. In particular, the small angle neutrons scattering (SANS) reduction workflows of interest. Lastly, conclusions and future work are presented in section~\ref{sec:conclusions} outlining the need for further co-design research efforts to provide optimal management strategies for the generated neutron sciences experimental data.

\section{Neutron Scattering Data Reduction Workflows} \label{sec:event_data}
\subsection{The NeXus file format} \label{subsec:NeXus}
SNS and HFIR instruments at ORNL use the international standard NeXus schema~\cite{Konnecke:po5029} for storing raw neutron event-based data. NeXus is based on the HDF5~\cite{hdf5} file format and follows a strict hierarchy for groups, datasets and attributes that identify each group of raw event based data from a neutron scattering experiment. Typical sizes for each file ranges between 0.1 up to 30 GB for each experiment depending on the complexity of the instrument and the number of entries of each dataset. These datasets are then stored in a ORNL-hosted warehouse in the neutron science computing facilities,~\url{analysis.sns.gov},  which already amounted to 1.6\,PB of available experimental raw data as of 2020.

The NeXus schema is illustrated in Table~\ref{tab:nexus} for the metadata structure saved to an underlying HDF5 file. Each level in the hierarchy maps to a ``group" in the underlying HDF5 dataset that is described with a string attribute with key ``NX\_class" to identify the group type according to the data source of information. Two representative groups are shown for: i) logs, NX\_class=NXlog, and ii) bank event data entries, NX\_class=NXevent\_data, which represent the majority of the processed group data type as described in subsection~\ref{subsec:Mantid}. Log entries are essentially process variables stored as time-stamped data which serve as a link to raw event data entries. Actual value entries, such as arrays or single values, are represented as scientific datasets (SDS) entries, or NX\_class=SDS in the NeXus schema. Thus, SDS entries don't require explicit attribute annotation in the NeXus stored metadata on-disk as they map directly to the HDF5 definition of a dataset~\cite{hdf5}. As a result, NeXus event-based datasets have a hierarchical metadata structure in which group types are the first level searching criteria.

\begin{table}[ht]
\begin{center}
\begin{tabular}{ l l }
Data Type  & Entry Name \\
\hline
group      & /entry \\
attribute  & /entry/NX\_class \\ 
           & ...                \\
group      & /entry/DASlogs \\
attribute  & /entry/DASlogs/NX\_class $\rightarrow$ ``NXlog" \\
group      & /entry/DASlogs/BL6:CS:DataType \\
attribute  & /entry/DASlogs/BL6:CS:DataType/NX\_class \\
dataset (SDS)    & /entry/DASlogs/BL6:CS:DataType/average\_value \\
dataset (SDS)   & /entry/DASlogs/BL6:CS:DataType/average\_value\_error \\
           & ... \\
group      & /entry/bank1\_events \\
attribute  & /entry/bank1\_events/NX\_class $\rightarrow$ ``NXevent\_data" \\
dataset (SDS)   & /entry/bank1\_events/event\_id \\
dataset (SDS)    & /entry/bank1\_events/event\_index \\
           & ... \\
group      & /entry/bank91\_events \\
attribute  & /entry/bank91\_events/NX\_class $\rightarrow$ ``NXevent\_data" \\
dataset (SDS)   & /entry/bank91\_events/event\_id \\
dataset (SDS)    & /entry/bank91\_events/event\_index
\end{tabular}
\end{center}
\caption{Schematic representation of the hierarchical NeXus schema~\cite{Konnecke:po5029} for recorded raw event-based neutron data.}
\label{tab:nexus}
\end{table}

\subsection{Data Reduction Workflows on Mantid}\label{subsec:Mantid}
\begin{figure}[!t]
\centering
\includegraphics[width=3.3in]{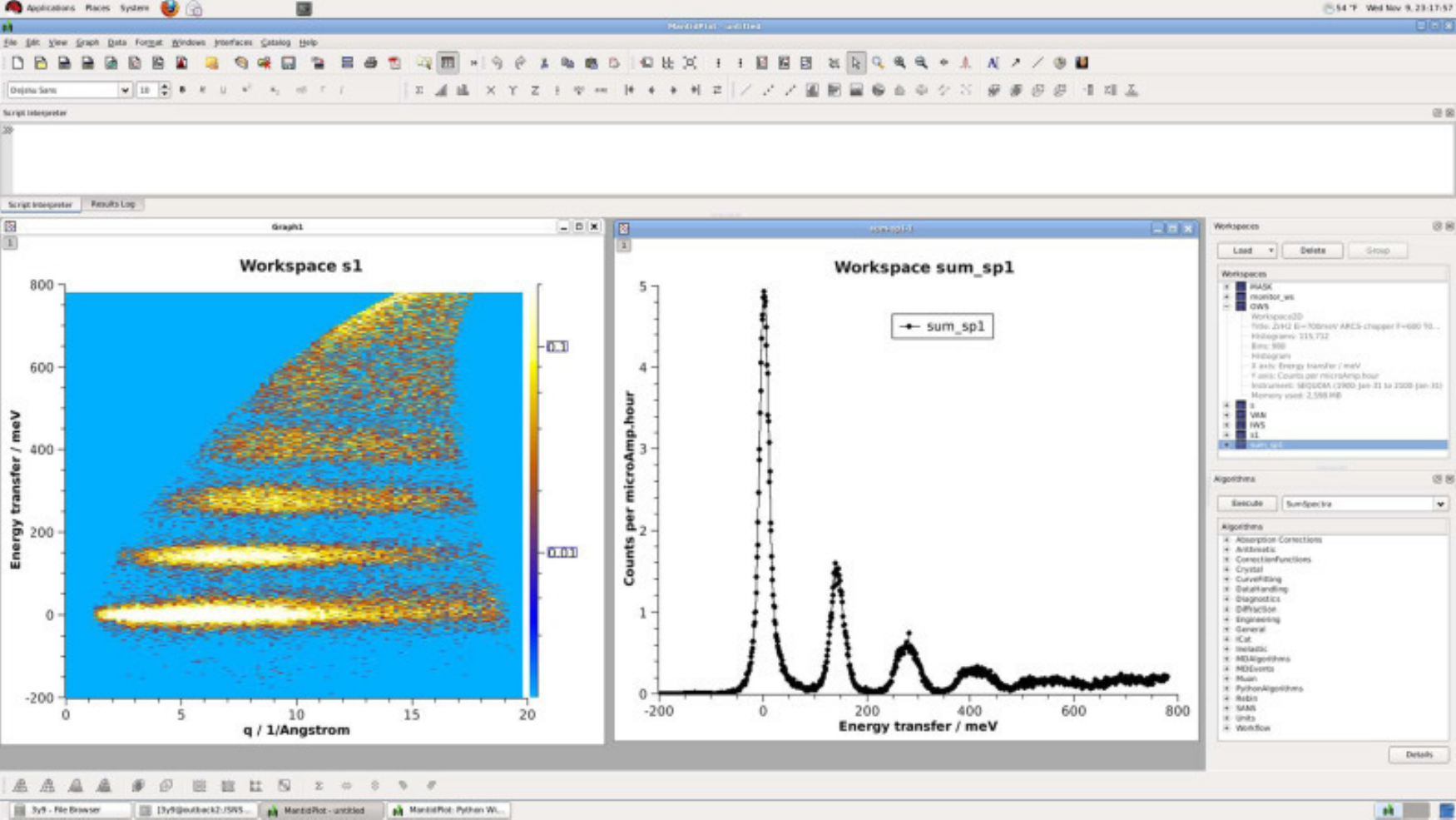}
\caption{Mantid graphical interface illustrating a reduced ``EventWorkspace" and physical quantities generated from NeXus files, from~\cite{ARNOLD2014156}.}
\label{fig:MantidGUI}
\end{figure}

NeXus files are processed in data reduction workflows using the Mantid framework~\cite{ARNOLD2014156}. These workflows include several input NeXus files for the physical interpretation, analysis and visualization tasks required by users of SNS and HFIR instruments. Figure~\ref{fig:MantidGUI} shows the typical user interactions of typical reduction workflows through the Mantid interface, in which several MB or GB of NeXus data is reduced to a histogram or a pixelated image. Reduction workflows call a single and unified ``LoadEventNexus" Mantid function for each NeXus file. ``LoadEventNexus" return an in-memory Mantid structure called an ``EventWorkspace" which is designed specifically for sorting time-of-flight event histograms~\cite{PETERSON201524}. To build a reduced ``EventWorkspace", ``LoadEventNexus" requires internal calls to different algorithms processing different parts of the NeXus file entries. Particularly time consuming algorithms include those processing logs and bank event data and those forming the in-memory metadata index at each step for each group. For more details, the reader is referred to the documentation of the ``LoadEventNexus" algorithm on Mantid~\cite{LoadEventNexus}.

\begin{figure}[!t]
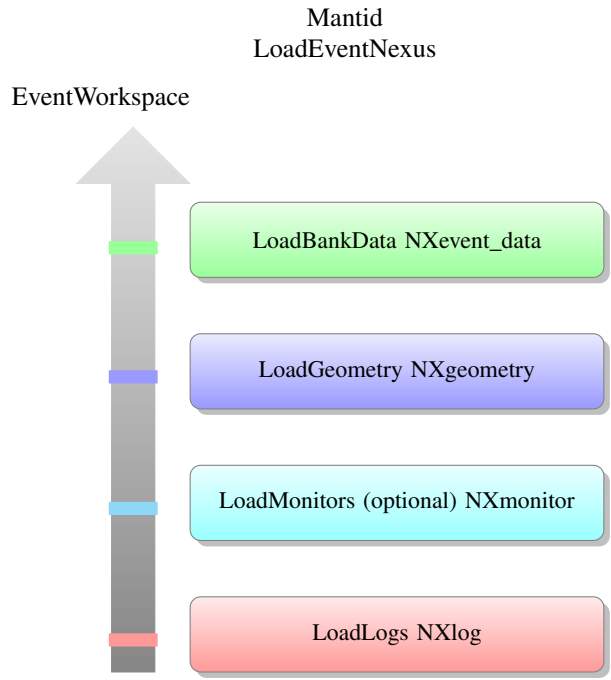

\begin{center}
\smartdiagramset{priority arrow width=1.5cm,
priority arrow height advance=2.0cm,
priority arrow head extend=0.2cm}
  Mantid \\ 
  LoadEventNexus
\begin{flushleft}
EventWorkspace
\end{flushleft}
\smartdiagram[priority descriptive diagram]{LoadLogs NXlog,LoadMonitors (optional) NXmonitor,LoadGeometry NXgeometry,LoadBankData NXevent\_data}

\begin{flushleft}
Raw input data \textless instrument\textgreater\_\textless\,run\_number\textgreater.nxs.h5
\end{flushleft}
\end{center}
\caption{Mantid's LoadEventNexus steps for processing entries of a single input NeXus file identified by instrument and experiment run number in data reduction workflows.}
\label{fig:LoadEventNexus}
\end{figure}

Mantid's original architectural design isolates each algorithmic step shown in Fig.~\ref{fig:LoadEventNexus}, this encapsulation prevents sharing ``expensive" data resources, such as metadata index information and file handlers, among these steps. Therefore, the existing implementation reconstructs metadata index information for every group level that is accessed in the search for data entries. As a result, several extra calls are made to the underlying HDF5 library tracking the metadata in appropriate b-trees structures~\cite{hdf5}, as well as increasing the number of memory allocations required to reconstruct each hierarchical level index. The latter adds to the overall wall-time bottlenecks observed in several neutron data reduction workflows using Mantid.

\section{Proposed Data Management Strategy} \label{sec:proposed}
The present effort introduces new data management strategies in the stages of Mantid's ``LoadEventNexus" to address current I/O bottlenecks. First, an in-memory index binary tree structure is introduced along all the stages of ``LoadEventNexus". The index key consists of a string prefixed with the entry ``type", the value of the ``NX\_class" attribute for each entry in Table~\ref{tab:nexus}, followed by the absolute path to each entry. Second, the proposed index is generated as soon as a NeXus file is opened and reused in the stages of Mantid's ``LoadEventNexus". The intention is to replace the current I/O bottlenecks due to the cost associated with hierarchical metadata reconstruction at each NeXus group level, as those shown in Table~\ref{tab:nexus}. The goal is to also match the search and processing patterns of NeXus entries in ``LoadEventNexus", as illustrated in Fig~\ref{fig:LoadEventNexus}.  

\begin{figure}[!t]

NX\_class
\begin{center}
\begin{forest}
for tree={rectangle, minimum size=0.5cm, 
      edge={black, line width=1mm, shorten <=1pt, shorten >=1pt, -}}
[NXlog, fill=red!30
  [NXevent\_data,fill=green!30]
  [NXmonitor, fill=blue!30
    [SDS,fill=violet!30]
  ]
]
\end{forest}
\end{center}

NXlog
\begin{center}
\begin{forest}
for tree={rectangle, minimum size=0.2cm, fill=red!30,
      edge={black, line width=1mm, shorten <=1pt, shorten >=1pt, -}}
  [/entry/Log4 
    [/entry/Log3]
    [/entry/Log6 
      [/entry/Log5]
      [/entry/Log7]
    ]
  ]
\end{forest}
\end{center}

NXevent\_data
\begin{center}
\begin{forest}
for tree={rectangle, minimum size=0.2cm, fill=green!30,
      edge={black, line width=1mm, shorten <=1pt, shorten >=1pt, -}}
  [/entry/bank4\_events
    [/entry/bank3\_events]
    [/entry/bank6\_events 
      [/entry/bank5\_events]
      [/entry/bank7\_events]
    ]
  ]
\end{forest}
\end{center}

SDS: Scientific Dataset
\begin{center}
\begin{forest}
for tree={rectangle, minimum size=0.2cm, fill=violet!30,
      edge={black, line width=1mm, shorten <=1pt, shorten >=1pt, -}}
  [/entry/bank4\_events/..
    [/entry/bank3\_events/..]
    [/entry/bank6\_events/.. 
      [/entry/bank5\_events/..]
      [/entry/bank7\_events/..]
    ]
  ]
\end{forest}
\end{center}

\caption{Schematic representation of the efficient binary-tree in-memory index metadata for NeXus files entries classified by NX\_class types at the top level. Each NX\_class node (NXlog, NXevent\_data, SDS) is a binary-tree on its own.}
\label{fig:Index}
\end{figure}

\begin{table}[ht]
\begin{center}
\begin{tabular}{ l l }
Key: NX\_class  & Value: Sorted binary-tree with absolute-path entry key\\
\hline

NXcollection & /entry/DASlogs \\
             & /entry/DASlogs/BL6:CS:DataType/enum \\
             & /entry/DASlogs/BL6:Chop:Skf1:PhaseLocked/enum \\
             & /entry/DASlogs/BL6:Chop:Skf2:PhaseLocked/enum \\
             & ... \\ 
             & \\
NXdetector   & /entry/instrument/bank1 \\
             & /entry/instrument/bank2 \\
             & ... \\
             & /entry/instrument/bank48 \\
             &  \\
NXlog        & /entry/DASlogs/BL6:CS:DataType \\
             & /entry/DASlogs/BL6:CS:beamslit4 \\
             & /entry/DASlogs/BL6:Chop:Skf1:MotorSpeed \\
             & ... \\
             & \\
NXevent\_data & /entry/bank1\_events \\
              & /entry/bank2\_events   \\ 
              & ... \\
              &   /entry/bank48\_events  \\
              & \\
SDS &   /entry/DASlogs/BL6:CS:DataType/average\_value \\
    &   /entry/DASlogs/BL6:CS:DataType/average\_value\_error \\
    &   ... \\
    &   /entry/bank1\_events/event\_id \\
    &   /entry/bank1\_events/event\_index \\
    &   /entry/bank1\_events/event\_time\_offset \\
    &   /entry/bank1\_events/event\_time\_zero \\
    &   /entry/bank1\_events/event\_total\_counts \\
    &   ... \\
    &   /entry/bank48\_events/event\_id \\
    &   /entry/bank48\_events/event\_time\_offset \\
    &   /entry/bank48\_events/event\_time\_zero \\
    &   /entry/bank48\_events/event\_total\_counts \\
    &    ...
\end{tabular}
\end{center}
\caption{Resulting in-memory index implementation using C++'s \texttt{map<string,set<string>} data structure, showing the two search levels by i) NX\_class and ii) absolute-path entry for a single NeXus file. SDS entries are the largest sub-tree.}
\label{tab:IndexMemory}
\end{table}

Figure~\ref{fig:Index} and Table~\ref{tab:IndexMemory} show a schematic representation of the proposed index structure. The first search bucket of this binary tree is given by the number of entry types (NX\_classes), which is typically only a few groups in the NeXus file as described in~\ref{subsec:NeXus}. Each node is a binary-tree on its own, since the Scientific Dataset (SDS) type refers to actual values (single or array values) it is the node with the largest number of entries (NX\_class-entries). As a result, the complexity of a search for a given entry becomes logarithmic on the number of classes and the number of entries-per-class:

\begin{equation}
\text{search} \sim  \mathcal{O}\left(\log\left(\text{NX\_classes}\,\times\,\text{NX\_entries-per-class}\right)\right).
\end{equation}

The resulting index is immediately constructed in-memory and passed along the algorithms called inside ``LoadEventNexus" in Fig.~\ref{fig:LoadEventNexus}. The latter required architectural changes in the algorithms encapsulation inside Mantid to enable reusability of ``expensive" resources, such as the introduced binary tree index in Fig.~\ref{fig:Index}, to avoid frequent memory allocation operations.

From the implementation perspective, the available data structures from the C++ standard template library (STL)~\cite{10.5555/2543987} are used. The end result is a two-step ordered binary tree, ``\texttt{map<string,set<string>}", in which the STL ``map" and ``set" associative containers have logarithm complexity guarantees. Due to the average number of entries, typically in the lower thousands range (2,000-3,000), no added benefit was seen with other associative data structures, \textit{e.g.} hash. In addition, the sorted aspect is also desired to enable range loops as observed in the current implementation of ``LoadEventNexus"~\cite{LoadEventNexus}. For implementation details and, most importantly, results and performance reproducibility the reader is referred to the changes on Mantid's source code: \url{https://github.com/mantidproject/mantid/pull/28495} currently available on Mantid's latest development branch.

\section{Impact} \label{sec:results}

\subsection{Mantid LoadEventNexus Performance}
\begin{figure}[!t]
\centering
\subfloat[]{\includegraphics[width=3.5in]{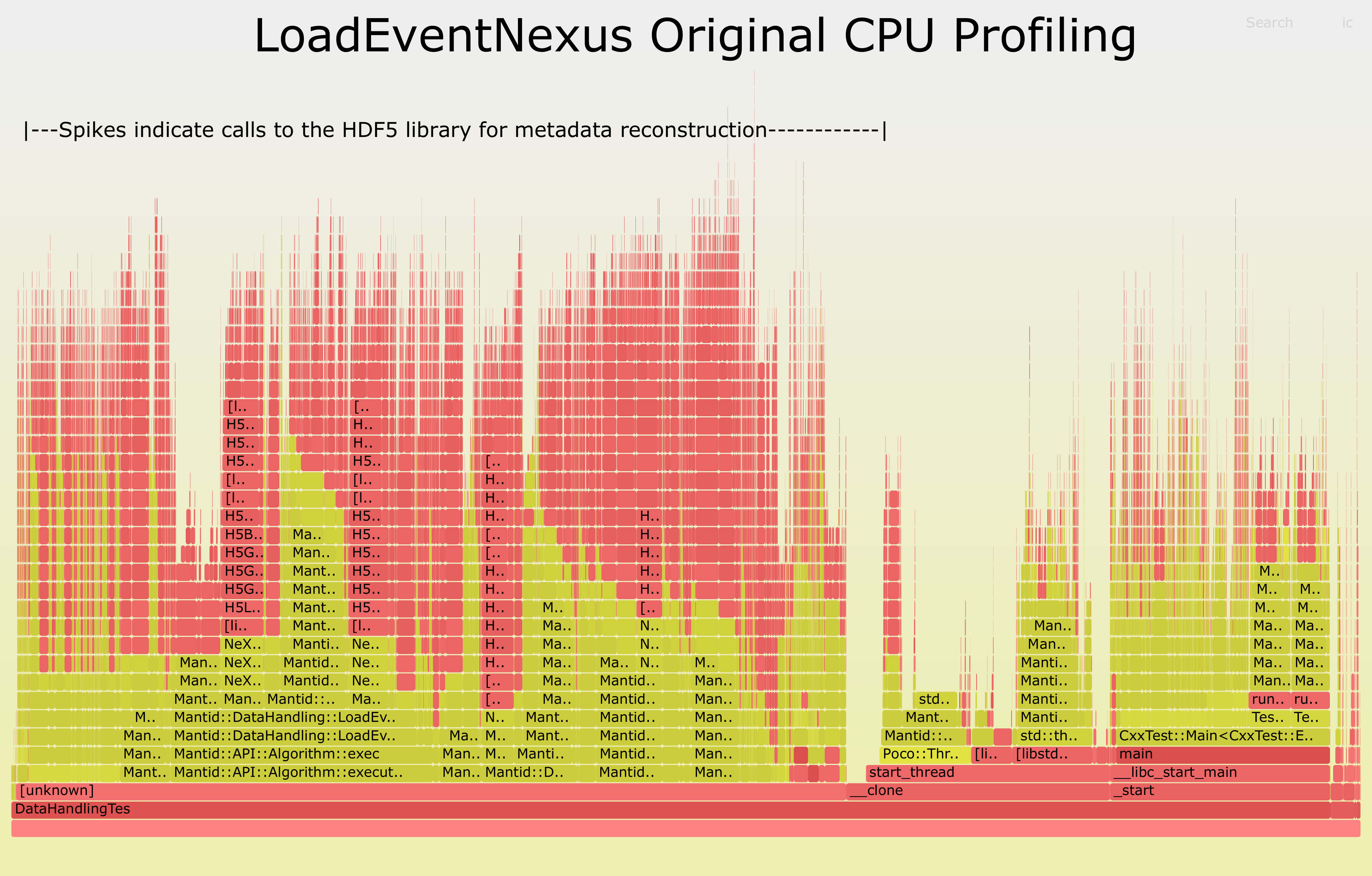}%
\label{fig:before}}
\hfil
\hfil
\subfloat[]{\includegraphics[width=3.5in]{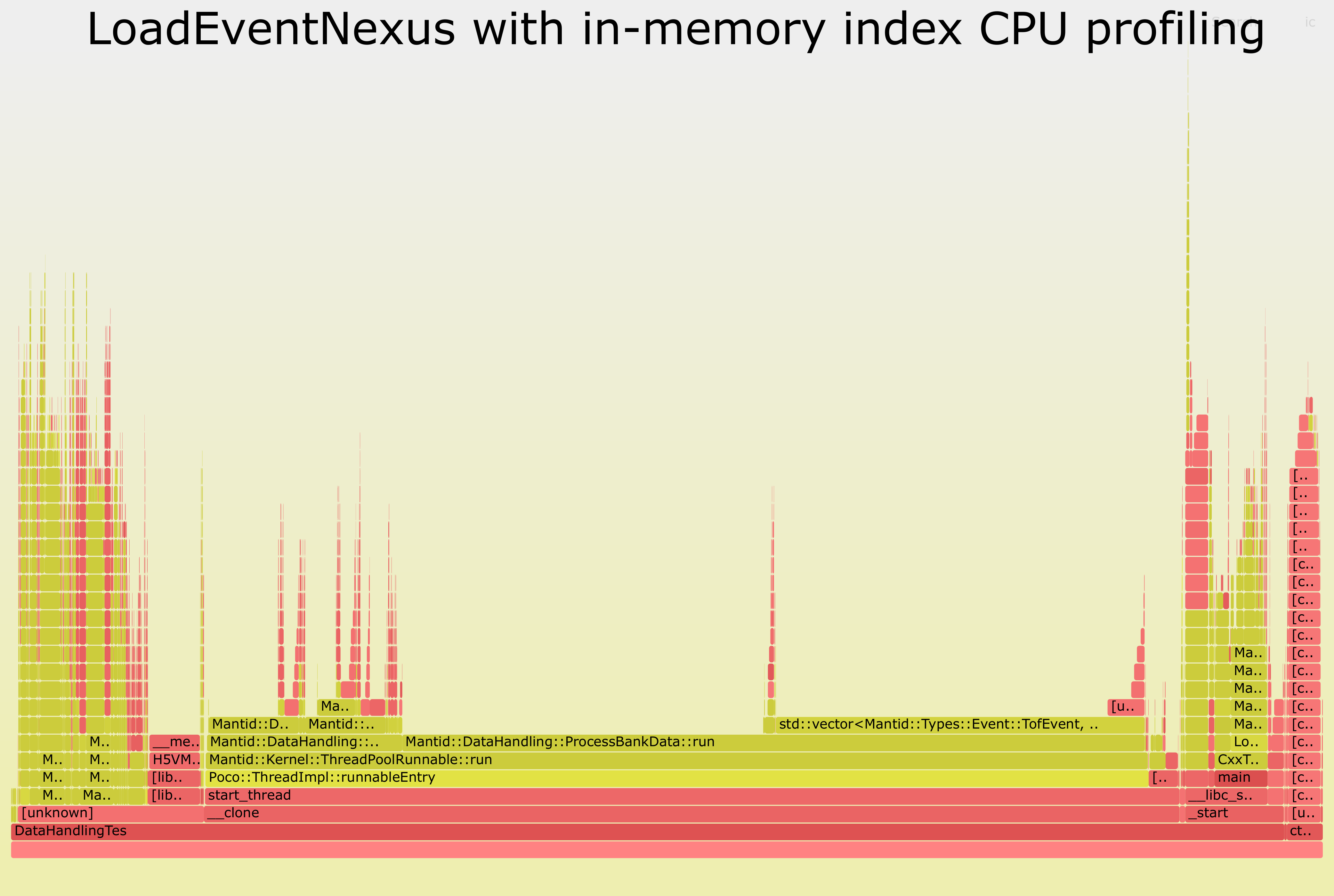}%
\label{fig:after}}
\caption{Mantid's ``LoadEventNexus" CPU profiling flame graph representation for (a) Mantid v5.0, (b) Mantid latest implementation with our proposed strategy. The reduction of metadata-related CPU operations bottlenecks is illustrated in this comparison.}
\label{fig:flamegraphs}
\end{figure}

The proposed changes in data management resulted in fewer metadata operations inside Mantid's ``LoadEventNexus". Figure~\ref{fig:flamegraphs} shows the flame graph~\cite{10.1145/2909476} representation (x-width illustrates the cost of each function, y-heigth is function call stack) of the CPU profiling for all the existing Mantid tests using ``LoadEvenNexus" for: (a) Mantid v5.0, and (b) Mantid latest development branch with the proposed improvement. It can be seen that the CPU time spent on tasks related to metadata management for entry search have been largely reduced for a variety of files.

Further measurements are provided to understand the impact on each individual NeXus file generated from different SNS and HFIR instruments at ORNL neutron facilities. Each instrument generates a set of ``runs", with each run stored as a NeXus file. Selected files are provided in Table~\ref{tab:runs} for a variety of instruments highlighting the different number of NeXus entries and sizes, which are typically processed with Mantid's ``LoadEventNexus".

\begin{table}[!t]
\begin{center}
\begin{tabular}{ l | l | r  }
\hline
NeXus file &	Entries &	Size (MB) \\
\hline
CG2\_8179 (GPSANS) &	3,683 &	62 \\
CG2\_8947  &	3,712 &	725 \\
CG3\_943  (BIOSANS) & 3,203	& 71 \\
CG3\_816   & 3,203	& 766 \\
CG3\_1545  &	3,607 &	137 \\
CG3\_1056 &	3,387 &	269 \\
CG3\_1003 &	3,387 &	1800 \\
CORELLI\_83353 & 2,660 &	297 \\ 
CORELLI\_145950 & 2,974 &	510 \\
EQSANS\_112300 & 2,529 & 461 \\
EQSANS\_113407 & 2,532 & 5800 \\
NOM\_78093 & 1,572 & 1100 \\
NOM\_78106 & 1,572 & 488 \\
\hline
\end{tabular}
\end{center}
\caption{Summary of selected representative NeXus files generated at ORNL instruments for Mantid's ``LoadEventNexus" performance comparison.}
\label{tab:runs_files}
\end{table}

\begin{figure}[!t]
\centering
\subfloat[]{\includegraphics[width=3.55in]{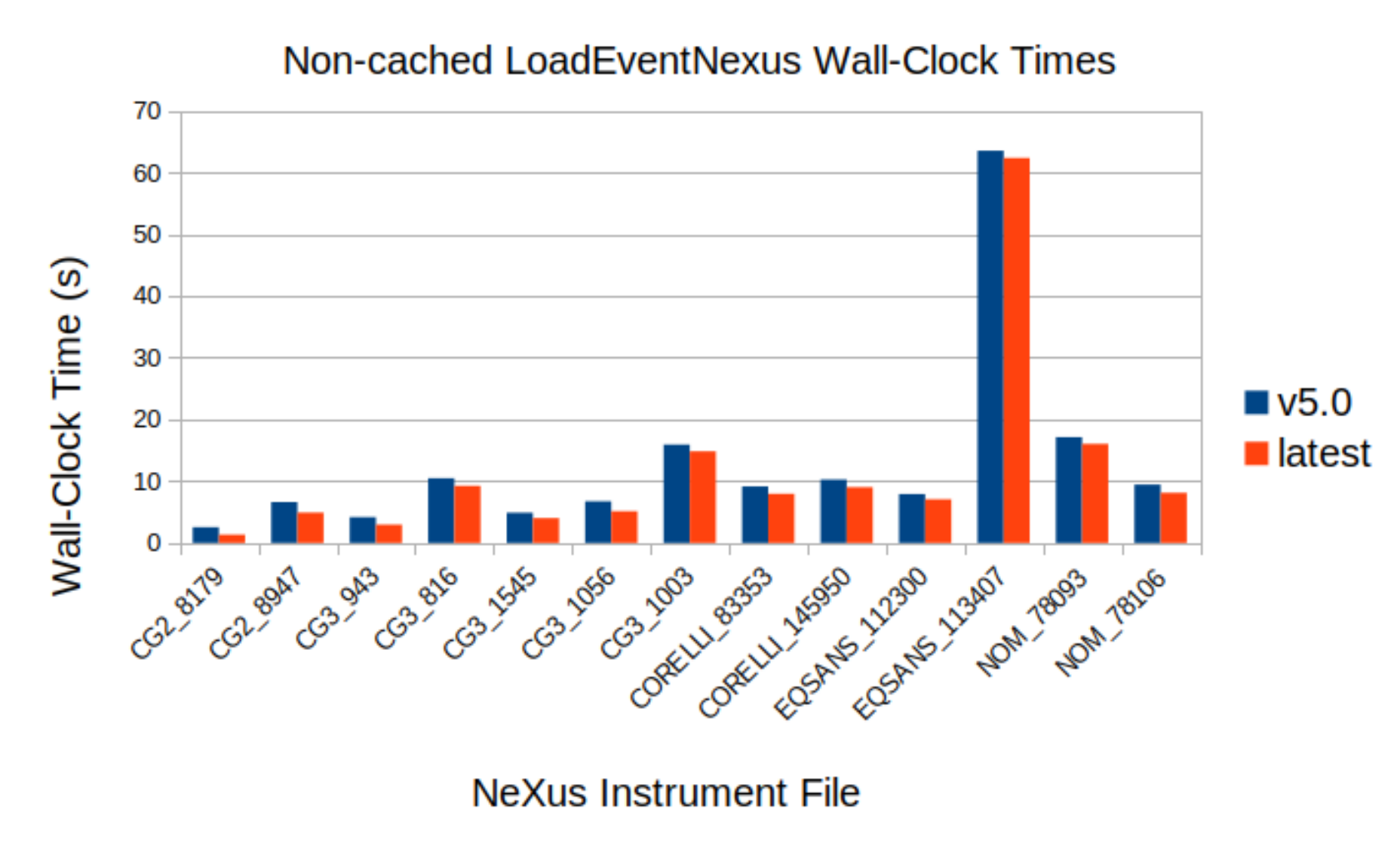}
\label{fig:cold_cache}}
\hfil
\hfil
\subfloat[]{\includegraphics[width=3.55in]{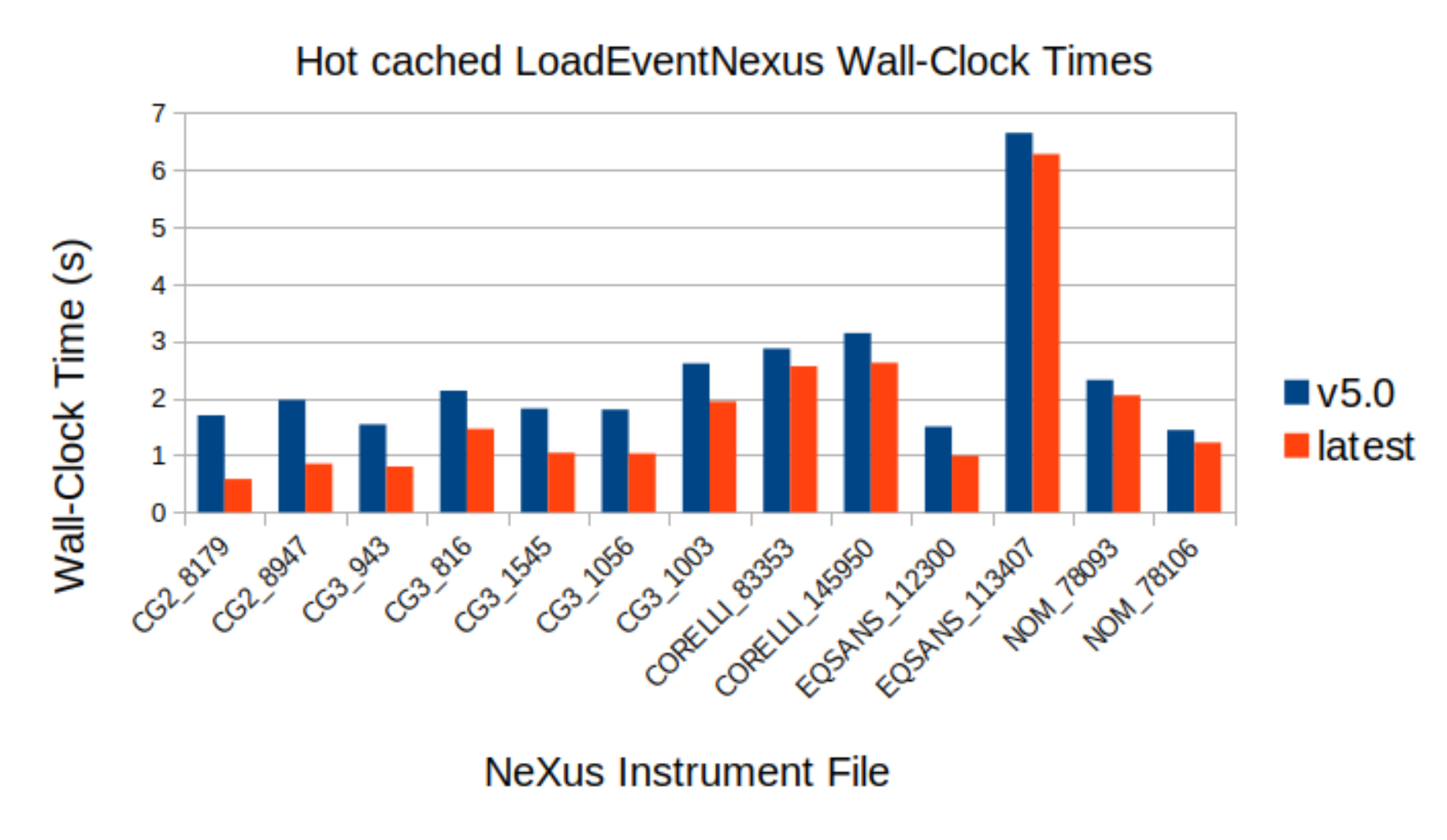}%
\label{fig:hot_cache}}
\caption{Comparison of Mantid's ``LoadEventNexus" wall-clock times for Mantid v5.0 release and our proposed strategy on Mantid's latest implementation. Results are shown for: (a) non-cached files (run once), (b) cached files showing varying improvements across different ORNL instrument generated files.}
\label{fig:performance}
\end{figure}

Performance of Mantid's ``LoadEventNexus" is measured for the v5.0 release version and compared against the latest development branch with the introduced changes from this work. As shown in Fig.~\ref{fig:performance}, wall-clock times are reported for ``LoadEventNexus" on the NeXus files listed on Table~\ref{tab:runs_files} running on a AMD Ryzen 7 3700X 8-Core processor, 64 GB of RAM, and a Hitachi HDP72505 500 GB hard drive for file storage. For completeness, we provide measured wall-clock times using ``non-cached" files (not previously used), and ``hot cached" files (previously used) to cover the different scenarios in which a NeXus file could be retrieved by users. Overall, results in Fig.~\ref{fig:performance} demonstrate that the changes introduced from this work provide a consistent speed up across different instrument generated NeXus files. Impact may vary depending on the file characteristics. For example, large files from EQSANS~\cite{Zhao:he5479} show little speed up in the wall-clock times, which indicates the need for identifying more bottlenecks in ``LoadEventNexus", while the smaller GPSANS (CG2)~\cite{BERRY2012179} files see larger benefits in speed up. More research is needed to understand the relationship between internal compute and I/O algorithms and file characteristics, in particular metadata entries and file sizes, as those shown in Table~\ref{tab:runs_files}. The long-term goal is to co-design efficient data reduction workflows as new use-cases are identified.

\subsection{ORNL Data Reduction Workflows}
Data reduction workflows are typically composed by a handful of NeXus files identified by different runs, such as those presented in Table~\ref{tab:runs_files}. The end result is the reduction of the raw NeXus event-based data into physical quantities of interest, in particular histograms and images. The current algorithmic improvements are then applied to three data reduction workflows of interest for small angle neutron scattering (SANS) instruments at ORNL facilities.

Table~\ref{tab:runs} shows results for the SANS instruments reduction workflows running on ORNL production systems at \url{analysis.sns.gov}. The production hardware consist of an Intel Xeon CPU E5-2670 v3 with 48 cores equipped with 512 GB of RAM. The SANS instruments consist of one time-of-flight instrument, EQSANS~\cite{Zhao:he5479}, where information from each event is used to reduce the data; and two monochromatic instruments, BIOSANS~\cite{Heller:kk5171} and GPSANS~\cite{BERRY2012179}, where event data is traditionally not used. Relative speed ups are presented as the ratio of the difference between wall-clock times obtained with Mantid v5.0 and the latest development version with the current algorithm improvements divided by the original wall-clock time obtained with Mantid v5.0.

These data reduction workflows are composed of different NeXus files, each representing a run number, thus impact might vary according to the I/O and computation characteristics ratio from different calls to ``LoadEventNexus". Overall, it can be seen that improvements apply consistently to all the production workflows when wall-clock times are measured before and after introducing the proposed index structure. As expected from our initial single file assessment for ``LoadEventNexus" in Fig.~\ref{fig:performance}, the GPSANS data reduction workflows shows the largest improvements, 30\% speed ups, which are typically composed of a large number of entries and small file sizes. On the other hand, improvements on EQSANS data reduction workflows reach a reproducible 10\% speed up, which are a more modest improvement as expected from the results in Fig.~\ref{fig:performance}. BIOSANS improvements are placed in between at 19\%, even though the workflow takes the longest as more data process is required. Overall, we proved that the improvement are universal and impact a wide range of NeXus files and their composition in data reduction workflows.

\begin{table}[ht]
\begin{center}
\begin{tabular}{ l l r r r l }
ORNL       & NeXus   & Max       & Mantid  &  Mantid & Relative \\ 
Instrument & entries & file size     & v5.0 WC  &  latest WC &  speed  \\ 
Workflow   & approx  & (MB) & time (s) &  time (s)  &
up \\
\hline
GPSANS~\cite{BERRY2012179} &  3,700 & 45  & 58.9           & 41.8          & 29\% \\
BIOSANS~\cite{Heller:kk5171}  &  3,700 & 444 & 100.2           & 80.9          & 19\% \\
EQSANS~\cite{Zhao:he5479}  &  2,500 & 62  & 99.0           & 88.0          & 11\% \\
\hline
\end{tabular}
\end{center}
\caption{Overall wall-clock (WC) times comparison and speed up on production data reduction workflows for SNS/HFIR instruments running on \url{analysis.sns.gov} hardware system.}
\label{tab:runs}
\end{table}

\section{Conclusions} \label{sec:conclusions}
This work introduces efficient data management strategies to address I/O bottlenecks in existing data reduction workflows at ORNL neutron scattering experimental facilities. For reproduciblity, the present work is available in the latest development branch of the Mantid data analysis and visualization framework. These improvements are also expected to benefit the larger Mantid community at other neutron source facilities around the world, such as ISIS (UK) and ILL (France), as they impact several NeXus files with a wide range of entries and file sizes. Efficient metadata indexing search is introduced using an entry ``absolute path" key binary-tree, while reduction of CPU runtime and memory footprint is achieved by modifying the current encapsulation in Mantid algorithms that process NeXus files. The overall impact on wall-clock time results in speed ups ranging from 11\% to nearly 30\% in current data reduction workflows of interest for Small Angle Neutron Scattering (SANS) instruments at ORNL. Future direction includes continuing researching different data management strategies to further customize existing reduction workflows. The latter is expected to focus on specific areas such as: event data filtering, histograms generation, data storage compression, and machine learning applications.

\section*{Acknowledgment}
Work at Oak Ridge National Laboratory was sponsored by the Division of Scientific User Facilities, Office of Basic Energy Sciences, US Department of Energy, under Contract no. DE-AC05-00OR22725 with UT-Battelle, LLC. We would like thank Dr. Mathieu Doucet, Dr James Kohl, and Mr. Rich Crompton of the Neutron Sciences Division at Oak Ridge National Laboratory for their helpful input to this work.

\ifCLASSOPTIONcaptionsoff
  \newpage
\fi

\bibliographystyle{IEEEtran}
\bibliography{IEEEabrv,paper.bib}

\end{document}